%% file: arxiv.tex
\documentclass[aps,pra,showpacs,amssymb,nofootinbib,superscriptaddress,twocolumn,longbibliography]{revtex4-1}
\usepackage[T1]{fontenc}
\usepackage[english]{babel}
\usepackage[utf8]{inputenc}
\usepackage{amsmath,amssymb}
\usepackage[colorlinks=true,citecolor=blue]{hyperref}
\usepackage{graphicx}
\usepackage{amsmath}
\usepackage{slashed}
\usepackage[table,xcdraw,dvipsnames]{xcolor}
\usepackage{bbm}
\usepackage{soul}
\usepackage{array,booktabs}
\usepackage{natbib}
\usepackage{amsmath}
\usepackage{bm}
\usepackage{bbold}
\usepackage{hhline}
\usepackage{xcolor}
\usepackage{mathtools} 
\usepackage{float}
\usepackage{tabularx}
\usepackage{tikz}

\usepackage{MnSymbol}

\newcommand{\bmsection}[1]{\par\medskip\noindent \textbf{#1 - }}

\newcommand{\tr}{\textnormal{Tr}}

\newcommand{\arxivFig}[1]{#1}
\newcommand{\osaFig}[1]{}
\newcommand{\AddS}[1]{\textcolor{black}{#1}}

\newcommand{\AddB}[1]{\textcolor{black}{#1}}
\newcommand{\AddC}[1]{\textcolor{black}{#1}}

\begin{document}

\title{Referenceless characterisation of complex media using physics-informed neural networks}

\author{Suraj Goel}
\email{gs74@hw.ac.uk}
\affiliation{Institute of Photonics and Quantum Sciences, Heriot-Watt University, Edinburgh EH14 4AS, UK}

\author{Claudio Conti}
\affiliation{Department of Physics, University Sapienza, Piazzale Aldo Moro 2, 00185 Rome, Italy}
\affiliation{Institute for Complex Systems, National Research Council (ISC-CNR), Via dei Taurini 19, 00185 Rome, Italy}
\affiliation{Research Center Enrico Fermi, Via Panisperna 89a, 00184 Rome, Italy\\}

\author{Saroch Leedumrongwatthanakun}
\affiliation{Institute of Photonics and Quantum Sciences, Heriot-Watt University, Edinburgh EH14 4AS, UK}

\author{Mehul Malik}
\email{m.malik@hw.ac.uk}
\affiliation{Institute of Photonics and Quantum Sciences, Heriot-Watt University, Edinburgh EH14 4AS, UK}


\begin{abstract}

\input{abstract}

\end{abstract}


\maketitle

\input{body}

\input{backmatter}

\bibliography{references}

\end{document}

%% file: abstract.tex
In this work, we present a method to characterise the transmission matrices of complex scattering media using a physics-informed, multi-plane neural network (MPNN) without the requirement of a known optical reference field. We use this method to accurately measure the transmission matrix of a commercial multi-mode fiber without the problems of output-phase ambiguity and dark spots, leading to upto 58\% improvement in focusing efficiency compared with phase-stepping holography. We demonstrate how our method is significantly more noise-robust than phase-stepping holography and show how it can be generalised to characterise a cascade of transmission matrices, allowing one to control the propagation of light between independent scattering media. This work presents an essential tool for accurate light control through complex media, with applications ranging from classical optical networks, biomedical imaging, to quantum information processing.

%% file: body.tex
\section{Introduction}
A scattering matrix provides complete knowledge of the linear optical response of a material. This not only gives us a better understanding of light-matter interactions, but also allows us to harness their use in a diverse range of photonic technologies~\cite{Kim2015,rotter_light_2017,Wetzstein2020,Cao2022}. For example, knowledge of the scattering matrix can be utilised in the control of light propagation through an opaque scattering medium~\cite{vellekoop_focusing_2007,popoff_measuring_2010}, serve as an alternative way to construct optical devices~\cite{Cao2022}, and has many applications spanning from imaging through scattering tissue~\cite{Choi2012,Katz2022}, mode-multiplexing through a multimode fiber~\cite{carpenter_degenerate_2012}, optical neural networks~\cite{Farhat1985,lin_all-optical_2018,Ugur2021,Logan2022} to the transport and manipulation of quantum states of light~\cite{Loffler2011,valencia_unscrambling_2020,Hugo2016,leedumrongwatthanakun_programmable_2020,goel_inverse-design_2022}. However, as the complexity of an optical system of interest grows larger, the efficient and accurate measurement of its scattering matrix can be challenging.

Over the last few decades, the development of techniques for measuring a scattering matrix--- both its transmission (TM) and reflection (RM) components---have been advanced and tailored to particular scenarios with different conditions and prior assumptions~\cite{Peres1989,popoff_measuring_2010,di_leonardo_hologram_2011,Rahimi-Keshari2012,Laing2012,Choi2013,Gerardin2014a,Yu2015,Tillmann2016,Dhand2016,Spagnolo2017,gordon_coherent_2019,Gordon2019,Lorenzo2022}. 
Conventionally, the measurement of a TM is performed under the assumption that a medium preserves the coherence properties of the input light, i.e., the scattering process (channel) is pure. 
In this case, the measurement is usually performed by sending different pure input states in a given basis sequentially to probe a medium/system of interest and detecting the corresponding output fields by means of off-axis holography using an external reference~\cite{Yamaguchi1997,Cuche2000,cizmar_shaping_2011,Ploschner2015}. 
The technique requires an interferometer that is insensitive to environmental disturbances, particularly for a long optical path length.


\arxivFig{\input{figure_tex/mpnn_model}}

To mitigate the problem of stability, alternative methods such as phase-stepping holography with a co-propagating internal reference field have been developed~\cite{popoff_measuring_2010}. Nevertheless, the use of a common-path reference field for an accurate TM measurement poses additional challenges since the internal reference needs to interfere with every optical mode within the complex medium with sufficient interference visibility for an optimal TM measurement. However, the internal reference field usually turns into a speckle pattern due to scattering, with a large variance in amplitude and consequently interference visibility, leading to a drawback known as the ``dark-spot problem''~\cite{Popoff2010a,cizmar_shaping_2011,hofer_manipulating_2019}. 
An alternative way to characterise complex media without using a reference field has been achieved through optimisation algorithms, which search for complex vectors using intensity-only measurements under the assumption of a pure process matrix. Various phase retrieval techniques have been developed, such as ones using Bayesian approaches~\cite{dremeau_reference-less_2015}, generalized approximate message passing~\cite{Schniter2014}, alternating projection methods such as the Gerchberg-Saxton algorithm\cite{Huang2021,Ancora2021}, and convex relaxations~\cite{NGom2016,ngom_mode_2018,suess_rapid_2020}.


A caveat of all these techniques that do not involve an external reference is that they do not completely characterise the transmission matrix. This is due to the fact that the intensity-only measurements used in these techniques cause phase ambiguity at the output, resulting in the relative phases between output modes being undefined~\cite{popoff_controlling_2011}. While a TM obtained in this manner is sufficient for the majority of imaging applications through complex media~\cite{Popoff2010a,caramazza_transmission_2019}, complete coherent information about the TM is necessary for many applications such as programmable photonic circuits~\cite{matthes_optical_2019} and in quantum information processing, where complex media are used to transport~\cite{Loffler2011,valencia_unscrambling_2020} and manipulate quantum states of light~\cite{Hugo2016,leedumrongwatthanakun_programmable_2020,goel_inverse-design_2022}. An extension to phase-stepping holography allows us to characterise this output phase ambiguity by interfering with different optical modes after the scattering process\cite{carpenter_110x110_2014}. 
Alternatively, phase diversity techniques can be applied to characterise this phase ambiguity by effectively measuring the output field intensity at different propagation distances in free space~\cite{gordon_coherent_2019,gordon_quantitative_2019,gordon_full-field_2019}. However, these methods are still subject to the dark-spot problem mentioned above or require full reconstruction of the optical field.

In recent years, artificial neural networks such as perceptrons and convolutional neural networks have been applied for tasks such as image reconstruction, classification, and transfer through a scattering medium~\cite{Turpin2018,Borhani2018,Rahmani2018,Li2018a,Tang2022}, demonstrating their potential for learning the transfer function of a scattering medium. 
While light scattering through a complex medium is a linear process, its measurement in intensity is non-linear, which makes it a suitable system to model within the framework of artificial neural networks. Incorporating the physical model that describes this scattering process into a neural network architecture is thus a clear contender for solving optimisation problems~\cite{Karniadakis2021}. Unlike general machine-learning models, physics-informed neural networks are models based on physical systems, and thus do not require treating the algorithm like a black box, but indeed a simplification tool. Recent advances in optics-based machine learning have not only led us towards enhanced control through linear complex media\cite{caramazza_transmission_2019,li_compressively_2021, matthes_learning_2021}, but also useful applications in non-linear microscopy~\cite{darco2022} and programming quantum gates~\cite{marcucci_programming_2020}.

In this work, we demonstrate the complete characterization of a coherent transmission matrix without the use of a reference field and the accompanying problem of dark spots by employing the use of physics-informed neural networks, referred to as a multi-plane neural network (MPNN). We do so by performing a set of random measurements before and after the complex medium, while probing only intensity at the output and subsequently feeding the data into a neural network designed to mimic the experimental apparatus. We demonstrate the improved accuracy of our technique by comparing the focusing efficiency achieved through the medium using the transmission matrices obtained with an MPNN to ones recovered with conventional phase-stepping (PS) holography. Furthermore, we investigate the number of measurements required to characterise the TM and show that while phase-stepping requires fewer measurements, the TM recovered with MPNN is much more accurate. We also show that our technique is significantly more noise-robust as compared to phase-stepping holography, allowing the recovery of a high fidelity TM in the presence of large amounts of noise. Finally, using a numerical simulation, we demonstrate the general usage of MPNNs for the TM measurement of a cascade of random scattering media. 

\section{Multi-Plane Neural Network}

\osaFig{\input{figure_tex/mpnn_model}}

We start by describing the model of a multi-layer optical network, where each complex medium is placed between reconfigurable phase-shifter planes, as illustrated in Fig.~\ref{fig:mpnn_model}. The $k$-th layer of the optical system is composed of a reconfigurable phase-shifter plane represented by a vector $x_k$ and a complex medium with complex transmission matrix $T_k$. The intensity $y$ observed at the output detectors of an $n$-layered network for a uniform excitation at an input of the network is given by
\begin{equation}
\begin{aligned} 
   y  &=  | \mathbf{F}\mathbf{P_{n+1}}\mathbf{T_n}\mathbf{P_{n}} \dots \mathbf{T_k}\mathbf{P_{k}} \dots \mathbf{T_2}\mathbf{P_2}\mathbf{T_1}x_1|^2,\\
   &=  | \mathbf{F}(x_{n+1}\odot(\mathbf{T_n}(x_{n}\odot ...        \mathbf{T_2}(x_2  \odot \mathbf{T_1} x_1) ...)))|^2,\\
 \end{aligned} 
\label{eqn:MPNN_model}
\end{equation}
where $\odot$ represents an element-wise dot product, $\mathbf{T_k}$ is the transmission matrix of the $k^{\textrm{th}}$ complex medium in the network, $\mathbf{F}$ is a known complex matrix defined by the detection optics, and $\mathbf{P_{k}}=\text{diag}[x_k]$. Equation \ref{eqn:MPNN_model} also describes the neural network that models this optical process, where $\{x_i\}_i$ are $n$ input vectors, $\{\mathbf{T_i}\}_i$ are fully-connected ``hidden'' complex layers with a linear activation, and $\mathbf{F}$ is a fully-connected known complex layer
with a non-linear activation of $|.|^2$ to simulate intensity measurements. Since the input layers are located at different planes in the network, we refer to this model as the multi-plane neural network (MPNN). 


\input{figure_tex/concept}

We train the described model on a measured \AddS{randomized} dataset using Tensorflow and Keras on Python. Tensorflow provides an open-source framework that makes models like these convenient to implement and extend~\cite{abadi_tensorflow_2016}. Building models in this framework allowed us to use previous complex-value neural network layers developed in \cite{caramazza_transmission_2019} and makes our work open to extensions using different optimisation techniques. Our model is optimised using adaptive moment estimation, also referred to as the Adam optimizer \cite{kingma2014adam}, the mean-squared error loss of which is given by 
\begin{equation}
    MSE= \sum_i | \bar{y}_i -y_i |^2,
    \label{eqn:loss}
\end{equation}
where $i$ represents the index of a point in the dataset, $\bar{y}$ is the predicted output from the model, and $y$ is the measured output. Once the loss function is decided, gradients to each weight in the layer $\mathbf{T_k}$ are calculated with respect to the MSE using a chain-rule governed by the back-propagation algorithm\cite{rumelhart_learning_1986}. To ensure that the learning is efficient, we appropriately set the learning rate before beginning the optimisation while also reducing it during the optimisation if the loss plateaus. Post training, retrieving the weights of the layer $\mathbf{T_k}$ gives us the required transmission matrix of the $k^{\textrm{th}}$ complex medium.

\section{Experimental Method}
\label{sec:exp_method}


\input{figure_tex/learning_TM}

In this work, we use the formalism of MPNN to measure the coherent transmission matrix of a multi-mode fiber using two programmable phase planes. The phase planes are implemented on spatial light modulators (SLMs) placed before and after the MMF to probe the optical fields propagating through the fiber using intensity-only measurements. A schematic of the setup is illustrated in Fig.~\ref{fig:concept}a, where light from a superluminescent diode is filtered by a spectral filter centered at 810 nm (FWHM 3.1 nm) and coupled into a 2m-long graded-index multi-mode fiber (MMF, Thorlabs-M116L02) sandwiched between two SLMs (Hamamatsu LCOS-X10468). \AddB{ The requirement of two separate SLMs is not strict for these measurements and the setup can be designed such that a single SLM is employed with a double reflection before and after the MMF.} The MMF has a core size of 50 $\mu$m and supports approximately $200$ modes for each polarization at 810 nm wavelength. The telescopes placed between each component in the setup are designed such that the incident beam covers a large area on each SLM. In this particular experiment, we only control a single polarization channel of the MMF. However, the techniques discussed here can be equivalently applied to characterise both polarization channels simultaneously.

We choose to work in the so-called macro-pixel basis, which consists of groups of SLM pixels chosen such that the intensity per macro-pixel is approximately equal (Fig.~\ref{fig:concept}b). $\textrm{SLM}_{1}$ is used to prepare a set of input modes in the macro-pixel basis and $\textrm{SLM}_{2}$ in combination with a CMOS camera (XIMEA-xiC USB3.1) allows us to perform measurements on the output modes of the fiber. $\mathbf{T}$ denotes the optical transmission matrix between $\textrm{SLM}_{1}$ and $\textrm{SLM}_{2}$ in the macro-pixel basis. The field after $\textrm{SLM}_{2}$ is given by the element-wise product $x_2  \odot \mathbf{T} x_1$, where $x_{1(2)}$ is a vector corresponding to the hologram implemented on $\textrm{SLM}_{1(2)}$. Finally, this field is incident on the camera placed in the Fourier plane of $\textrm{SLM}_{2}$ using the transfer matrix $ \mathbf{F}$. The resultant measurement intensity $y$ on the camera is given by

\begin{equation}
    y=  | \mathbf{F} (x_2  \odot \mathbf{T} x_1)  |^2,
    \label{eqn:two_model}
\end{equation}
which describes a single-layered MPNN model. A set of random measurements are performed on the setup to generate a dataset for this model. A number of holograms are generated with phases randomly varied following a uniform distribution. These holograms are displayed on each SLM and the corresponding intensities measured by the CMOS camera are recorded. We sample the intensity of the field at the camera at positions separated by a distance corresponding to the size of a diffraction-limited spot (as shown in Fig.~\ref{fig:concept}b) calculated using the effective focal length of the lens system (L$_{9-11}$) between $\textrm{SLM}_{2}$ and the camera.

\input{figure_tex/dark_spot}

\AddB{While we expect the TM to have a dimension of approximately $200 \times 200$ based on fiber specifications, we oversample our measurements at the SLM planes to capture any optical misalignment and aberrations in the experiment~\cite{matthes_learning_2021}.} Thus, we use 800 macro-pixels at the input phase plane ($\textrm{SLM}_{1}$) and 832 macro-pixels at the output phase plane ($\textrm{SLM}_{2}$) to perform the measurements. \AddB{ In principle, any modal basis can be used to perform these measurements, however, we choose the macro-pixel basis because it provides accurate and low-loss phase-only modulation on the SLM.}  Since the intensity sampling at the camera plane is limited by the resolution of the optical system, we only use $367$ sampling points here.
We train our model containing an $800\times832$ dimensional TM. The optimization is carried out in multiple batches with a batch size of 500 samples, and is accelerated using a GPU (NVIDIA GeForce RTX 3060). We observe good convergence in about 20 epochs of training, as plotted in the training and test metrics shown in Fig.~\ref{fig:learning}(a-b). Retrieving the weights from the hidden layer of the model recovers the measured TM in the macro-pixel basis, as visualised in Fig.~\ref{fig:learning}c\AddC{, and can be shown using the supplemental codes and dataset (Ref.~\cite{Goel_Data_and_codes_2022})}.  Transforming this TM to the Laguerre-Gauss modal basis with 20 mode groups reveals a block-diagonal structure with crosstalk as shown in  Fig.~\ref{fig:learning}d, which is a typical structure expected for a graded-index multi-mode fiber~\cite{carpenter_110x110_2014,boonzajer_flaes_robustness_2018}.

\section{Results and Discussions}

\subsection{Accuracy of the measured transmission matrix \label{subsec:exp_Res}}

\input{figure_tex/FE_result}

 To verify the accuracy of the measured transmission matrix, we perform an optical phase conjugation (OPC) experiment to focus light from a given input mode into a particular output mode after scattering through the fibre, by displaying a phase-only solution on the first plane (SLM$_1$), the second plane (SLM$_2$), and both planes simultaneously. 
 The focusing efficiency obtained by controlling light using only the first plane or the second plane allows us to assess the quality of individual rows and columns of the transmission matrix, respectively. On the other hand, manipulating light using both planes allows us to assess the overall quality of the measured transmission matrix. The focusing efficiency is defined as the ratio of intensity measured in a diffraction-limited spot at the camera to that measured in an output region that is 1.75 times the area corresponding to the output facet of the fiber. This is to capture any light that is diffracted outside the output facet due to phase modulation at SLM$_2$.

We measure the TM of a multi-mode fiber using the MPNN technique and compare it with one measured using the phase-stepping (PS) technique. \AddS{It is important to note that the PS technique can be used to measure the full coherent TM without output-phase ambiguity via a two-step process as follows.} In the first step, we measure the joint transfer matrix of the fiber and $2f$-lens system ambiguous to the reference field, i.e. $ D~\mathbf{F}~\mathbf{T}$, where $D$ is an arbitrary diagonal matrix owing to the unknown reference field. We do so by displaying the superposition of each input mode with the chosen reference mode at $\textrm{SLM}_{1 }$ and varying their relative phase in multiple steps, while measuring the intensity at each output spot. \AddS{The output fields at a particular input mode are then reconstructed by using the Fourier-Transform reconstruction algorithm~\cite{chen2022phase,popoff_controlling_2011}.} In order to maximize the intensity per mode, we choose the input basis to be a discrete Fourier transform of the macro-pixel basis with the $50^{\text{th}}$ mode chosen as the reference. The second step involves the reconstruction of the input reference field $D$. To do so, we send the input reference mode using $\textrm{SLM}_{1}$ and perform the phase-stepping technique using $\textrm{SLM}_{2}$. We prepare the superpositions of each output mode with the chosen output reference mode on $\textrm{SLM}_{2}$ and vary their relative phase in multiple steps. The measured intensity at the camera thus allows us to reconstruct the matrix $D$ corresponding to the input reference field in a similar manner as the first step above. It should be noted that the knowledge of $D$ is unnecessary for many applications such as imaging through complex media as well as the aforementioned experiment on performing OPC using only the first plane. \AddS{All measured data with PS is processed using the Fourier transform reconstruction algorithm~\cite{bruning1974digital,surrel1996design}}.

\input{figure_tex/convergence}

In the case of performing OPC with only the first plane ($\textrm{SLM}_{1}$), the focusing efficiency achieved at different points across the output facet of the fiber using a TM obtained via the PS and MPNN techniques is shown in Fig.~\ref{fig:dark_spots}a. Using the PS technique, we observe a reduction of the focusing efficiency at several output points due to the dark-spot problem~\cite{cizmar_shaping_2011}. This is due to the nature of speckle that results in a high probability of obtaining very low output intensities of the internal reference mode after scattering through the fiber. This leads to low interference visibility in the PS technique for specific output modes, which consequently results in inaccuracy in the reconstructed transmission matrix at these outputs. This problem is solved by measuring the transmission matrix with the MPNN technique, as this does not involve a \AddS{static} reference mode \AddS{ but instead uses many random inputs to probe the scattering process}.
The more uniform focusing efficiency achieved across the output facet of the fiber with the MPNN technique is clearly illustrated in Fig.~\ref{fig:dark_spots}a.  A histogram of focusing efficiencies achieved with these two methods  (Fig.~\ref{fig:dark_spots}b) shows that low focusing efficiencies are only observed with the PS technique, while the MPNN technique achieves a significantly higher maximum focusing efficiency.


While improved control with the first phase plane using various optimisation techniques has been well studied in many previous works, one of the chief merits of our approach lies in the simultaneous measurement of relative phases between the rows of the transmission matrix, i.e. the coherence between output modes. We assess the accuracy of the reconstructed relative phase between output modes by performing an OPC experiment to focus light by using only the second phase plane ($\textrm{SLM}_{2}$). Log-scale images of a focused spot at the centre of the output facet of the fiber using the two techniques are compared in Fig.~\ref{fig:results_FE}a-b. We observe the suppression of unwanted speckle background when using a TM obtained with the MPNN technique as compared to the one acquired using the PS technique. By measuring the focusing efficiencies for different input modes, the overall enhancement obtained with the MPNN technique as compared to the PS technique is evident in Fig.~\ref{fig:results_FE}c. The average focusing efficiency using the second plane increases from $26.5 \pm 2.3 \%$ using PS to $40.8 \pm 1.7 \%$ using MPNN.  The underlying reason for this improvement is that learning the TM with the MPNN technique does not require a static internal reference mode, whereas the use of the fixed internal reference mode in the PS technique results in errors at particular outputs due to the dark-spot problem.


As discussed above, a complete characterization of the transmission matrix including relative phases between its rows and columns is essential for coherent control of optical fields propagating through a complex medium. To examine this coherent control, we perform an OPC experiment by focusing light at different output spots by simultaneously utilising both the phase planes at hand. The solution of phase patterns for focusing is determined using an iterative wavefront-matching technique \cite{sakamaki_new_2007,fontaine_design_2017}. As seen in the log-scale images shown in Fig.~\ref{fig:results_FE}d-e, light focused using both phase planes with a TM acquired using the MPNN method has significantly less speckle background as compared to one acquired with the phase-stepping technique. Quantitatively, we are able to achieve an average focusing efficiency of $65.5 \pm 2.5 \%$ (up to a maximum of $73.8 \%$) with both planes using the MPNN technique (Fig.~\ref{fig:results_FE}f). This is a substantial increase with respect to that achieved with the PS technique, where we observe an average focusing efficiency of $42.4 \pm 3.1 \%$ (up to a maximum of $46.7\%$). This result also demonstrates the increase in focusing efficiency achievable with two-plane control as compared to individual phase planes, which makes it particularly suitable for usage in applications such as programmable optical circuits~\cite{goel_inverse-design_2022,leedumrongwatthanakun_programmable_2020}. 

The maximum achievable focusing efficiencies  (ideal simulations) shown in Figs.~\ref{fig:dark_spots} and \ref{fig:results_FE} are numerically calculated by taking into account the presence of polarization coupling in the fiber. \AddB{The fiber TM is represented by a truncated random unitary matrix considering that only one polarization channel is measured and controlled. This results in the lowest achievable maximum focusing efficiency when only the first phase plane (SLM$_1$) is used for control. When only the second phase plane (SLM$_2$) is used, the maximum focusing efficiency is increased as compared to the first case. The third case entails both phase planes being used together to focus light through the MMF. As there is much more phase control achievable using the two phase planes, one can now focus light with much higher efficiency as compared to the previous two cases.}


\subsection{Efficiency of learning with the MPNN technique}

In this section, we study the number of measurements required to obtain an optimal TM using the MPNN technique as compared to the PS technique. First, we experimentally evaluate this by performing an OPC experiment with the fiber TM reconstructed with different dataset sizes. For the MPNN method, the size of the dataset ($\alpha$) is quantified by the total number of intensity-only measurements performed on the camera divided by the number of input modes that characterize the TM. For the PS method, this quantity is a bit more nuanced since we only control the number of phase steps ($n_{\phi}$) within the interval $[0, 2\pi]$ that are used per mode. As described in section~\ref{sec:exp_method}, the first part of the PS method requires $(n_{\text{in}}-1)n_{\phi} + 1$ measurements, where $n_{\text{in}}$ is the number of input modes and the additional measurement corresponds to that of the reference itself. The second part requires $(n_{\text{out}}-1)n_{\phi}  $ measurements where $n_{\text{out}}$ is the number of output modes measured on the camera. The total number of measurements for each input mode is then given by $\alpha \approx [(n_{\text{in}}+n_{\text{out}}) / n_{\text{in}}]n_{\phi}$. In our experiment, $n_{\text{in}}=800$ and $n_{\text{out}}=367$, which gives $\alpha \approx 1.46 n_{\phi}$ for the PS technique. In this manner, the parameter $\alpha$ corresponds to the total number of measurements performed per mode in both the MPNN and the PS technique.


\input{figure_tex/noise}

The focusing efficiency achieved with a TM obtained via the MPNN and PS techniques as a function of the number of measurements per mode ($\alpha$) is plotted in Fig.~\ref{fig:results_convergence} for all three cases---focusing with the first, the second, and both planes. For the PS technique, the focusing efficiency is seen to converge to its maximum value at $\alpha \sim 6-8$, which is close to the minimum required number of phase steps ($n_\phi \sim 3-4$)~\cite{Heinosaari2013}. However, it reaches a plateau after this and cannot be improved further \AddS{owing to the presence of noise in the experiment~\cite{Vellekoop2008a,Ylmaz2013,Jang2016a}.} In contrast, the focusing efficiency obtained via the MPNN technique is seen to converge at a higher number of measurements per mode ($\alpha=20$). However, in all three OPC cases, the maximum efficiency achieved with MPNN is higher than that achieved with the PS technique--- $46.5\%$, $43.6\%$, and $73.8\%$ versus $45.0\%$, $30.2\%$, $46.7\%$ with the first, second, and both planes respectively. In particular, the maximum efficiency is significantly higher when focusing with only the second or both phase planes---cases where complete coherent information of the TM plays a critical role. Thus, while the MPNN technique takes longer to learn a given TM, the reconstructed TM is more accurate, as quantified by the focusing efficiency achieved through it. One should note that the number of measurements can also be reduced by incorporating the underlying physical model of a multimode optical fiber~\cite{li_compressively_2021,matthes_learning_2021}.

\subsection{Noise-robustness of the MPNN technique}

From the previous sections, it is clear that one of the advantages of the MPNN technique is improved performance over the PS technique in the presence of noise. While our experiment studies one specific case, here we quantify this improvement by simulating the effects of different noise levels on both techniques. An $800\times800$-dimensional random unitary matrix is chosen as our ground truth TM and intensity measurements using the PS and MPNN techniques are simulated while varying the number of measurements per mode ($\alpha$). Noise in the measurement is modelled as additive white Gaussian noise on the readout intensity
\begin{equation}
    \tilde{y}_{\sigma}=  | \mathbf{F} (x_2  \odot \mathbf{T} x_1)  |^2 + \mathcal{N}(\AddS{\mu},\sigma),
    \label{eqn:noise_model}
\end{equation}

\noindent where $\mathcal{N}(\mu,\sigma)$ is additive \AddS{white} Gaussian noise with mean \AddS{$\mu=0$} and standard deviation $\sigma$. \AddS{ Such a noise model is standard and includes the effects of multiple random processes such as thermal noise~\cite{boncelet2009image, surrel1997additive}.}
The signal-to-noise ratio (SNR) is the ratio of the average norm of the signal intensity to the \AddS{noise standard deviation}, i.e. \AddS{SNR $= \overline{|y|}/\sigma $}~\cite{janesick2007photon}. 
It should be noted that each simulated data point is normalised to unity before adding noise, which simply implies an SNR of $1/\sigma$ in our model. The fidelity between the recovered TM ($\widetilde{\mathbf{T}}$) and the ground-truth TM ($\mathbf{T}$) is calculated as $\mathcal{F}(\mathbf{T})=|\tr(\widetilde{\mathbf{T}}^{\dagger}{\mathbf{T}})|^2/(\tr(\widetilde{\mathbf{T}}^{\dagger}\widetilde{\mathbf{T}})\tr(\mathbf{T}^{\dagger}{\mathbf{T}}))$. The fidelity is sensitive to the relative phases between rows and columns of the TM, and thus only reaches its maximum value of 1 when complete coherent information about the TM is present.

The TM fidelity as a function of the number of measurements ($\alpha$) is plotted in Fig.~\ref{fig:noise} for different levels of noise. In these simulations, we choose $n_{\text{in}}= n_{\text{out}} = 800$, which implies that $\alpha \approx 2 n_\phi$ for the PS technique. In the noiseless case (SNR = inf), the PS technique converges to perfect fidelity at $\alpha \approx 6$, while the MPNN technique requires $\alpha \approx 10$ to do the same. Note that in this case, the PS technique is able to reach perfect fidelity regardless of the dark spot problem, as even the smallest interference signal provides complete information with no noise present. As the SNR decreases, the maximum fidelity achievable via the PS technique drops rapidly and is unable to reach perfect fidelity regardless of the number of measurements used. For example, even with a small amount of noise (SNR=\AddS{20}), the PS technique can only recover a TM with fidelity less than 60\%. In contrast, the MPNN technique is able to achieve very high fidelity in the presence of noise. As can be seen in Fig.~\ref{fig:noise}b, we are able to recover a high-quality TM ($\mathcal{F}=80.87\%$) even when the SNR is as low as \AddS{0.8}, with fidelity gradually increasing with the number of measurements per mode ($\alpha$). This highlights the significantly improved noise-robustness of the MPNN technique in comparison to the PS technique.


\AddS{ 
Despite using the most noise-resilient algorithm to reconstruct the TM using PS data~\cite{chen2022phase}, the MPNN method significantly outperforms the PS technique. This is because the MPNN method uses a large part of the dataset together to minimize the defined loss function in Eq.~\ref{eqn:loss}. The addition of noise to the data merely adds local minima to the loss function across the optimization space, leaving the global minimum largely untouched. In contrast, PS relies on processing each data point individually over the $n_{\phi}$ phase-steps as a sinusoid, where the addition of noise can severely impact the visibility and phase reconstruction.}

\AddS{These results may give the impression that PS is an unreliable technique, however, this is untrue because here we have pushed the technique to its very limits. The SNR range where we perform our simulations is far below previously tested ranges with PS~\cite{ChenYujun2020, Jang2016a}, demonstrating a much superior noise resilience of MPNN. Moreover, fidelity is a very unforgiving metric as it requires both the relative phases and amplitudes of both rows and columns of the TM to be well reconstructed for it to be high. A poor quality TM recovered using PS with an SNR=5 and fidelity of 17\%, can still control light with first and second planes with about 17\% and 13\% focusing efficiencies respectively. Nonetheless, such a metric is critical in high-precision applications such as programmable quantum gates~\cite{goel_inverse-design_2022}, where slight errors in knowledge of the TM can drastically affect the performance of the entire experiment. }

In practice, our experimental results are still far from ideal as estimated by the numerical simulations (black dotted lines in Figs.~\ref{fig:dark_spots} and \ref{fig:results_FE}). This deviation from the ideal might originate from a variety of imperfections in the experiment that are not explained by the simple model of noise studied here. These include imperfections such as the choice of basis, instability of light source, phase instability of SLM, temperature-dependent movement of optomechanics and the optical medium, or linearity of the camera. Many of these issues can be addressed and improved in the experiment to achieve perfect focusing efficiency~\cite{gomes_near_2022,Mastiani2022,Turtaev2017}, and could be combined with the MPNN technique presented here. 


\subsection{Learning multiple transmission matrices with an MPNN}

\input{figure_tex/MPNN_res}

In this section, we demonstrate how the MPNN technique can be used to reconstruct a complex optical network characterised by a series of transmission matrices, as described in Eq.~\eqref{eqn:MPNN_model}. Furthermore, we show how knowledge of each individual TM allows us to focus light at any intermediate plane within this network. As shown in Fig.~\ref{fig:MPNN_results}a, we simulate a cascade of three $16\times16$-dimensional TMs ($\textrm{T}_i$) separated by programmable phase planes. These are followed by a known mode-mixer (F) that performs a 2D discrete Fourier transform. By randomly modulating the phase planes and performing intensity measurements after F, the MPNN technique is able to fully characterise this optical network. The recovered TM of each medium as well as the training loss are shown in Fig.~\ref{fig:MPNN_results}b-c. 

A multi-plane network such as this can allow us to not only control light through the whole system, but also control light at intermediate planes as conceptualised in Fig.~\ref{fig:MPNN_results}a. As an example, we simulate optical phase conjugation using the three recovered TMs to focus light at each intermediate plane in the system. Insets in Fig.~\ref{fig:MPNN_results}b show the focused image obtained at each plane in the trained network using the preceding phase planes. The size of the dataset required to characterise this network with dimension $3\times 16 \times 16$ is $\alpha \approx 10^5$  \AddC{as also shown in supplemental codes (Ref. \cite{Goel_Data_and_codes_2022}) }. However, this is not the minimal size required to train this model and strongly depends on the training parameters. Further tuning and investigations can lead to a better understanding of how the minimum dataset size required to train this model varies with the number of planes and dimensions of the TM. Nonethless, it should be noted that the model for such a neural network is very complex and to our knowledge, MPNN is the only known method to date that can perform such a task.

A caveat of our technique is that the measured TM of two consecutive complex media in the series can be ambiguous up to a diagonal matrix on either side. In a cascade of complex media $T_i$ separated by planes $P_i$, taking a single layer at the $k^{\text{th}}$ plane,
\begin{equation}
    T_k P_k T_{k-1} =    T_k D P_k D^{-1} T_{k-1} =    \widetilde{T}_k P_k \widetilde{T}_{k-1}
    \label{eqn:mpnn_ambig}
\end{equation}
where $D$ is a diagonal matrix. Due to the commutation relation between $D, P_k$ and $D^{-1}$, the MPNN technique measures the TM of an individual complex medium up to such an ambiguity in the TMs of consecutive media, for example, the presence of equal but opposite diagonal phases. We anticipate that this ambiguity affects the training since few elements in $D$ can acquire high amplitudes leading to over-fitting, however, this problem can be tackled using suitable regularizers. Importantly, this ambiguity between $T_k$ and $\widetilde{T}_k$ does not affect the description of the overall cascaded optical network. 

Systems consisting of cascaded programmable phase planes separated by optical media are fast gaining popularity, with recent work demonstrating their use in a variety of applications such as spatial mode-sorting \cite{fontaine_design_2017,fontaine_laguerre-gaussian_2019,fontaine2021hermite}, projective measurements \cite{hiekkamaki2019near}, unambiguous state discrimination \cite{goel2022simultaneously}, programmable quantum circuits \cite{goel_inverse-design_2022, goel2023unveiling,hiekkamaki_high-dimensional_2021}, and optical neural networks \cite{lin_all-optical_2018,wang2021diffractive}. In all these implementations, accurate knowledge of the optical system between phase planes is critical to their performance. While free-space propagation is relatively straightforward to model, aberrations arising from the devices and elements used can introduce significant design challenges. By enabling the characterisation of a cascade of independent TMs, the MPNN technique provides a way to more accurately design such multi-plane devices, with uses ranging from classical to quantum optical networks.

\section{Conclusion}
In this work, we have conceptualised and experimentally demonstrated a method to characterise a complex medium, or a network of complex media, using physics-informed neural networks, referred to as multi-plane neural networks (MPNN). We apply the proposed MPNN technique to measure the full coherent transmission matrix of a multi-mode fiber without the need for an external reference field. The key idea behind the measurement is to randomly modulate phases of optical fields both before and after the fiber and measure the intensity-only outcomes to form a dataset. The trained model produces a transmission matrix capable of controlling light by manipulating fields not only before the multi-mode fiber, but also after it, which relies on the complete coherent information of the obtained transmission matrix without the problem of dark spots and output-phase ambiguity. We demonstrate accurate control of optical modes through a multi-mode fiber using the MPNN method, with a significant improvement over the phase-stepping technique in terms of focusing efficiency and noise-robustness. Finally, we show the capability of this technique to learn more complex systems such as a cascade of transmission matrices interspersed with multiple phase planes and discuss possible applications. Our technique will allow for accurate control of coherent light not only through complex media but also through complex optical networks, with applications ranging from optical communication systems to biomedical imaging.

%% file: figure_tex/mpnn_model.tex
\arxivFig{\begin{figure*}[t!]\centering\includegraphics[width=0.8\textwidth]{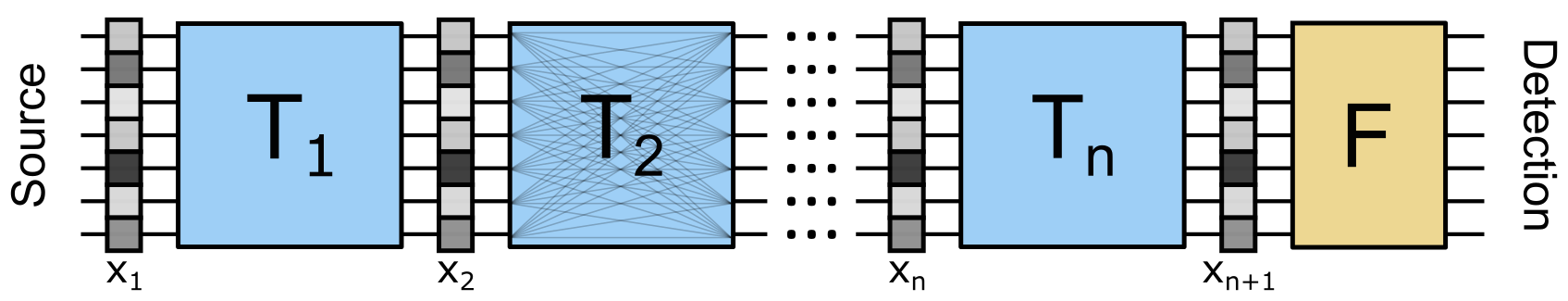}}
\osaFig{\begin{figure}[t!]\centering\includegraphics[width=\textwidth]{figures/mpnn_model.png}}

\caption{Schematic of an $n$-layered optical network: a cascade of $n$ complex media denoted by their optical transmission matrices $T_k$ are separated by reconfigurable phase-shifter planes $x_k$, followed by detection optics $F$. This optical network can be represented as a multi-plane neural network (MPNN) with $n+1$ input layers of $x_k$, $n$ ``hidden-layers'' for $T_k$, and a known layer for $F$ with a $|.|^2$ activation.}
\label{fig:mpnn_model}

\arxivFig{\end{figure*}}
\osaFig{\end{figure}}

%% file: figure_tex/concept.tex
\arxivFig{\begin{figure*}[t!]}
\osaFig{\begin{figure}[t!]}

\centering\includegraphics[width=\textwidth]{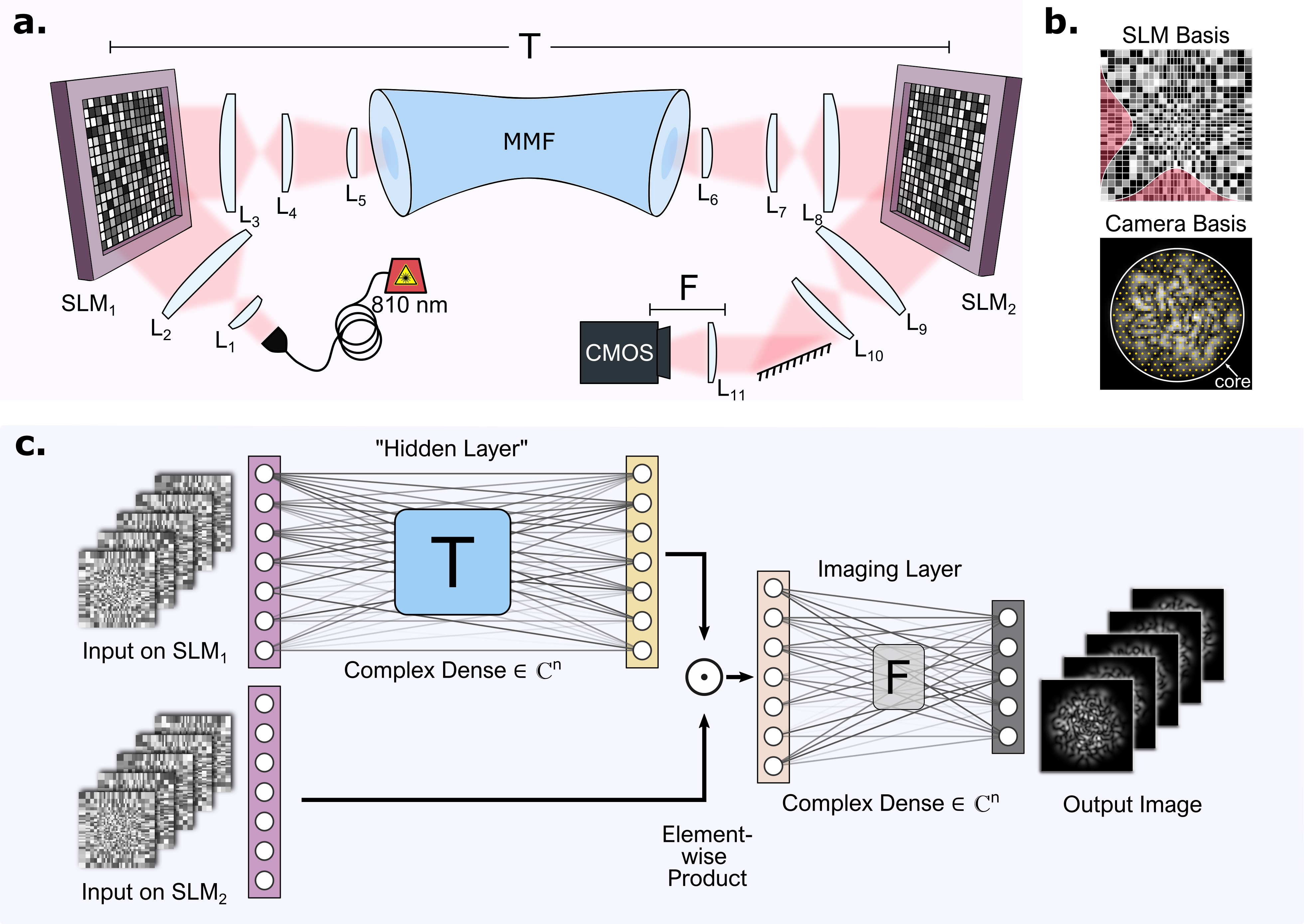}
\caption{
(a) Experiment: Light from a superluminescent diode is filtered by a 3.1nm spectral filter centered at 810 nm, modulated by a random phase pattern displayed on a spatial light modulator (SLM$_1$) and coupled into a multi-mode fiber (MMF) which is the complex medium of interest. The output speckle field from the MMF is projected onto SLM$_2$, which displays another random phase pattern, and is then detected on a CMOS camera which is placed in the Fourier plane of SLM$_2$. (b) The phase patterns on the SLM are constructed in a specific macro-pixel basis with varying pixel size based on the incident field distribution. Intensities of the output modes are recorded at a given set of points at the camera, which enclose an area corresponding to the MMF core. (c) The physics-informed neural network consists of two input layers encoding phase patterns on SLM$_1$ and SLM$_2$, and a single output layer encoding the intensity pattern of the output speckle in the given basis. The hidden layer $T$ denotes the complex transformation between SLM$_1$ and SLM$_2$ in the macro-pixel bases, while $F$ is a known layer corresponding to the $2f$-lens system between SLM$_2$ and the camera.}
\label{fig:concept}

\arxivFig{\end{figure*}}
\osaFig{\end{figure}}

%% file: figure_tex/learning_TM.tex
\arxivFig{\begin{figure*}[t!]}
\osaFig{\begin{figure}[t!]}

\centering\includegraphics[width=\textwidth]{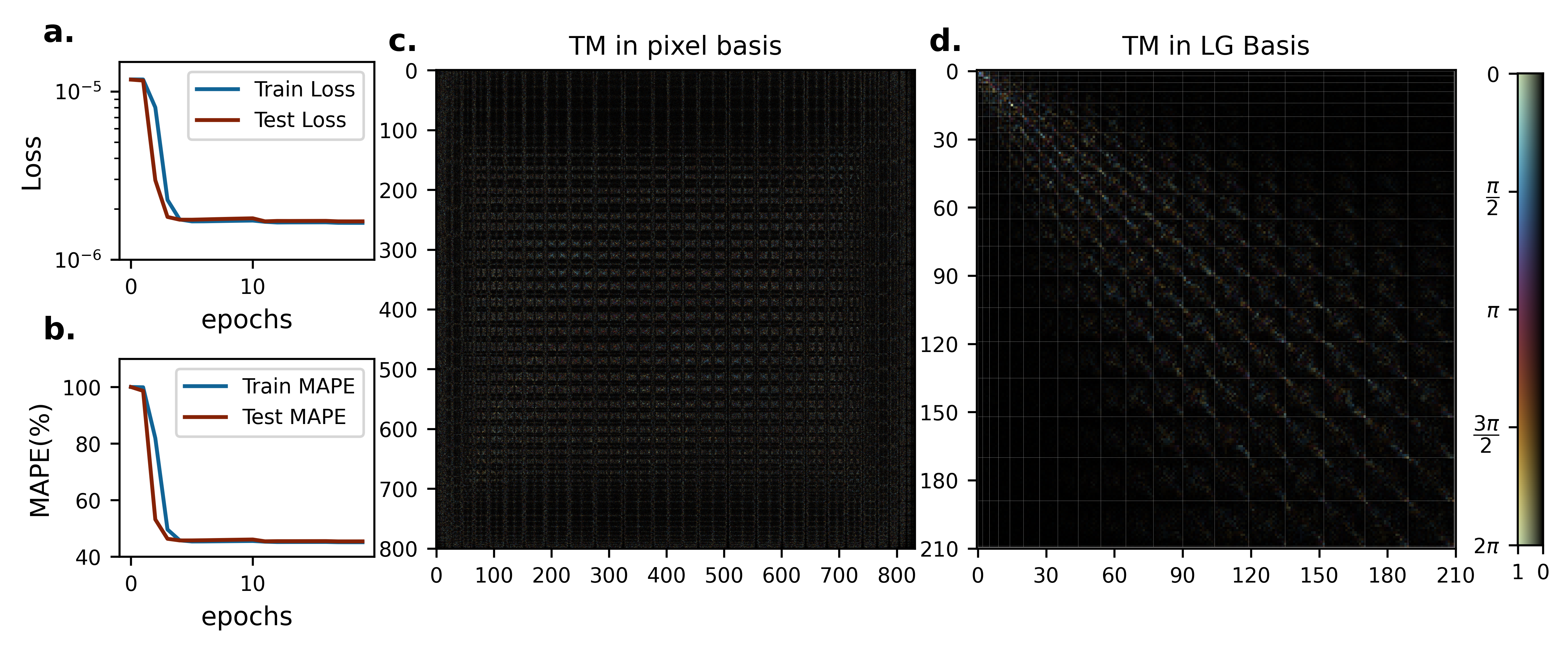}
\caption{Learning the coherent transmission matrix of a multi-mode fiber: Test and train (a) loss and (b) mean-absolute percentage error (MAPE) as learning proceeds in epochs. A visual plot of the learnt transmission matrix in the (c) macro-pixel and (d) Laguerre-Gauss (LG) basis, where modes are segregated into respective mode groups. In order to change the basis of the TM from (c) to (d), a transformation matrix $B_{\textrm{Pix} \rightarrow \textrm{LG}}$ is constructed that maps the set of macro-pixel modes to a set of LG modes. The TM in the LG basis is calculated as $T_{\textrm{LG}} = B_{\textrm{Pix} \rightarrow \textrm{LG}}^\dagger T_{\textrm{Pix}} B_{\textrm{Pix} \rightarrow \textrm{LG}}$.}
\label{fig:learning}

\arxivFig{\end{figure*}}
\osaFig{\end{figure}}

%% file: figure_tex/dark_spot.tex
\arxivFig{\begin{figure*}[t!]}
\osaFig{\begin{figure}[t!]}

\centering\includegraphics[width=\textwidth]{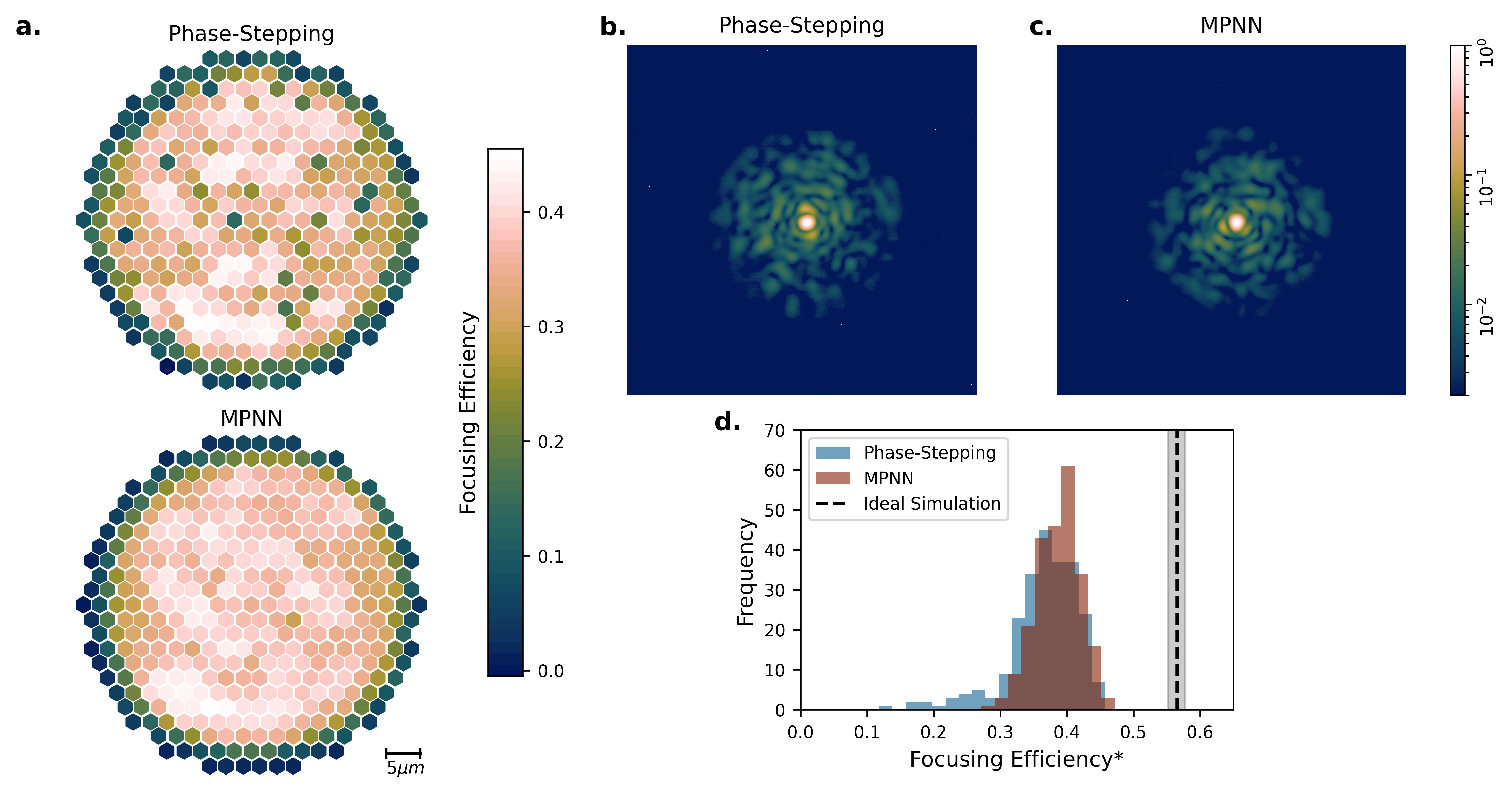}
\caption{A comparison of optical phase conjugation (OPC) performed with only the first phase plane: (a) Focusing efficiencies achieved at different points across the output facet of the multi-mode fiber using the phase-stepping (PS) and multi-plane neural network (MPNN) techniques. Output points with low focusing efficiency due to the inherent dark-spot problem can be seen in the PS method (see text for more details). In contrast, focusing efficiencies obtained with the MPNN method show more uniformity across the fiber facet. \AddB{ Log-scale images of light focused using the first SLM with a TM obtained by the (b) PS and (c) MPNN techniques. (d)} A histogram comparing the PS and MPNN methods shows that the PS technique leads to some output points with very low focusing efficiencies, while those obtained with MPNN method are more uniform and significantly higher. Note: Only the points within first 80\% diameter of the core from the center of the core are taken for the purpose of this histogram.}
\label{fig:dark_spots}
\arxivFig{\end{figure*}}
\osaFig{\end{figure}}

%% file: figure_tex/FE_result.tex

\arxivFig{\begin{figure*}[t!]}
\osaFig{\begin{figure}[t!]}

\centering\includegraphics[width=\textwidth]{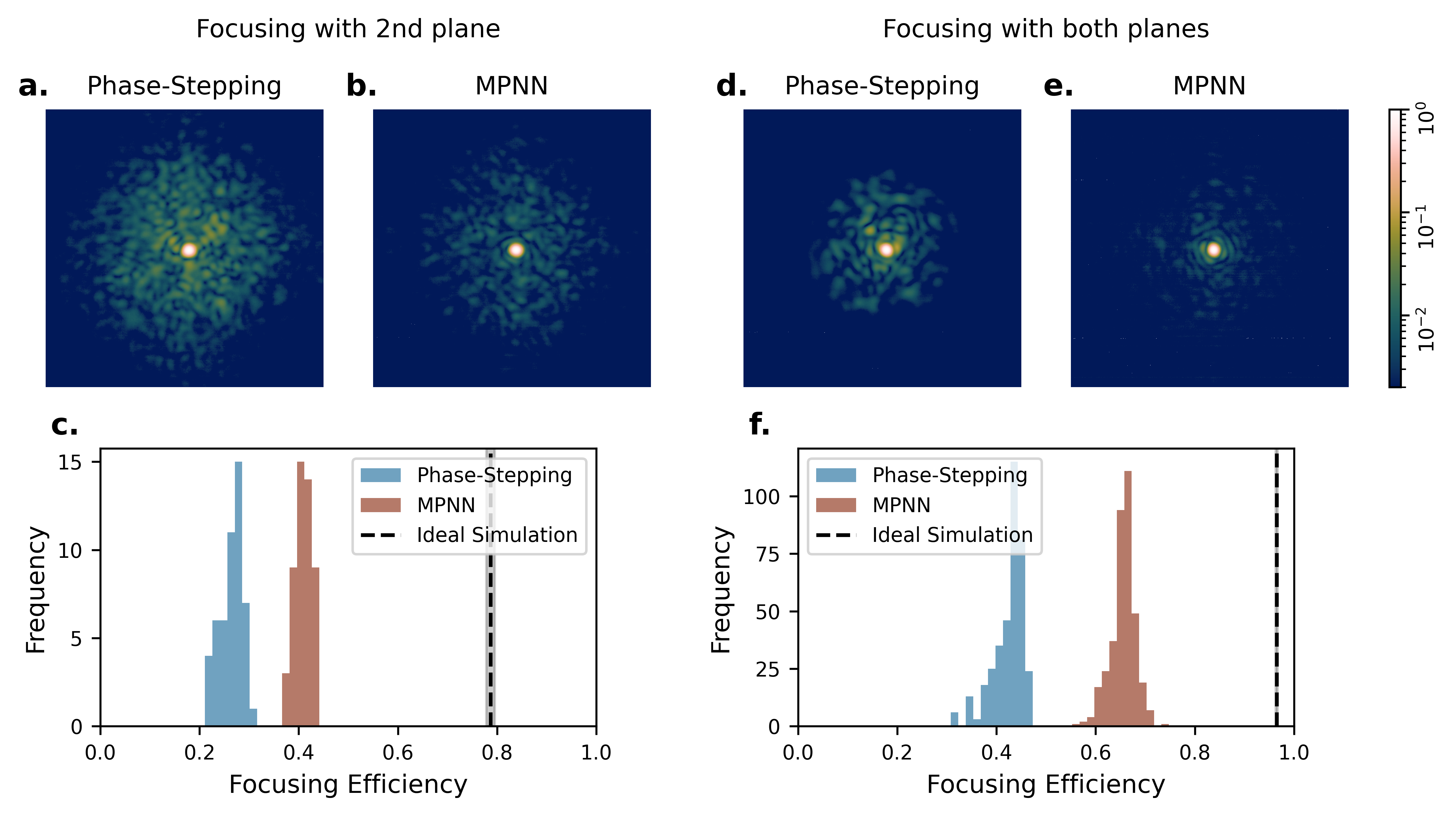}
\caption{(a-c) A comparison of optical phase conjugation (OPC) performed with only the second phase plane. Log-scale images of light focused using only $\textrm{SLM}_{2}$ with a TM obtained by the (a) phase-stepping and (b) multi-plane neural network techniques. (c) A histogram comparing second-plane focusing efficiencies of 50 random input modes achieved with the PS and MPNN techniques shows a marked improvement with the latter. (d-f) A comparison of OPC performed with both phase planes. Log-scale images of light focused using both SLMs with a TM obtained by the (d) PS and (e) MPNN techniques. (f) A histogram comparing focusing efficiencies for all output modes achieved with both planes shows a significant improvement with MPNN over the PS technique. The focusing efficiencies are corrected for SLM$_2$ basis-dependent loss for both methods.}
\label{fig:results_FE}

\arxivFig{\end{figure*}}
\osaFig{\end{figure}}

%% file: figure_tex/convergence.tex
\arxivFig{\begin{figure*}[t!]}
\osaFig{\begin{figure}[t!]}

\centering\includegraphics[width=\textwidth]{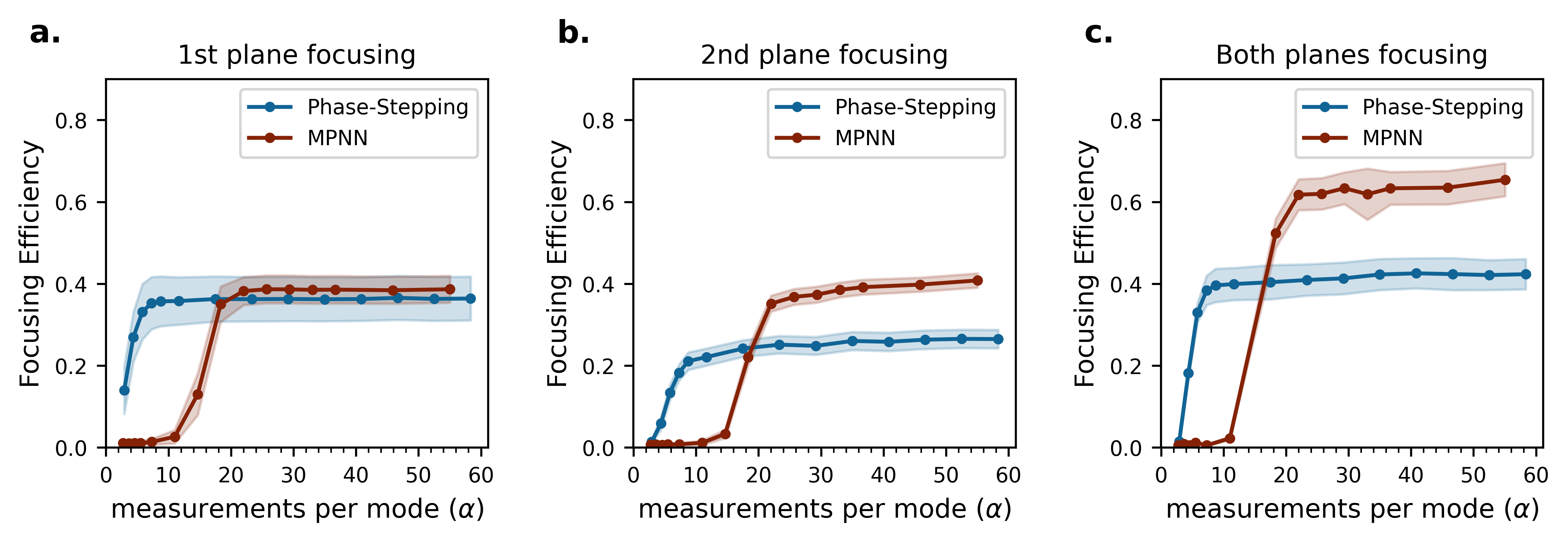}
\caption{Number of measurements required for optimal transmission matrix measurement: Experimental focusing efficiency achieved with (a) the first phase plane, (b) the second phase plane, and (c) both planes simultaneously, using a TM reconstructed with the phase-stepping technique and the multi-plane neural network (MPNN) plotted as a function of the number of measurements per input mode ($\alpha$). While phase-stepping converges to a maximum faster, the MPNN technique shows a much higher focusing efficiency.}
 \label{fig:results_convergence}

\arxivFig{\end{figure*}}
\osaFig{\end{figure}}

%% file: figure_tex/noise.tex

\arxivFig{\begin{figure*}[t!]}
\osaFig{\begin{figure}[t!]}

\centering\includegraphics[width=\textwidth]{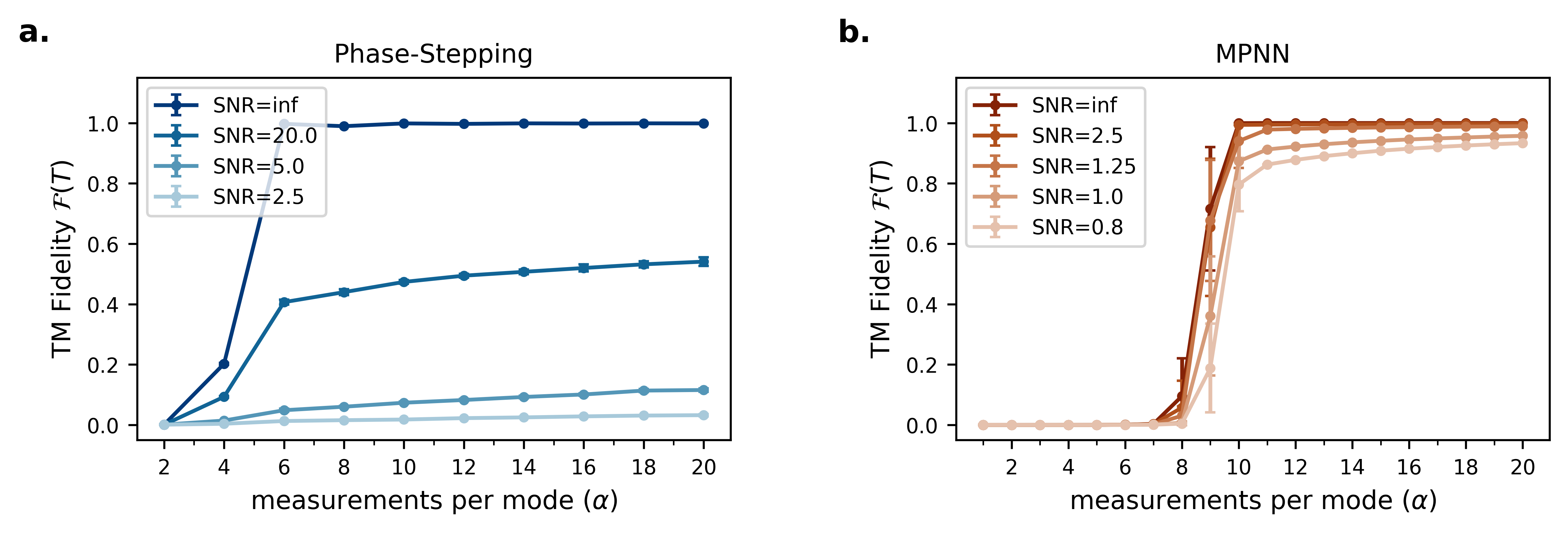}
\caption{Noise-robustness of MPNN: A simulated $800 \times 800$- dimensional TM is recovered using the (a) phase-stepping (PS) and (b) multi-plane neural network (MPNN) techniques for different additive gaussian noise levels. The fidelity of the recovered TM is plotted as a function of the number of measurements per input mode ($\alpha$). While the PS technique quickly degrades in the presence of noise, the MPNN technique is able to reach high fidelities with small increases in the number of measurements.}
 \label{fig:noise}

\arxivFig{\end{figure*}}
\osaFig{\end{figure}}

%% file: figure_tex/MPNN_res.tex

\arxivFig{\begin{figure*}[t!]}
\osaFig{\begin{figure}[t!]}

\centering\includegraphics[width=\textwidth]{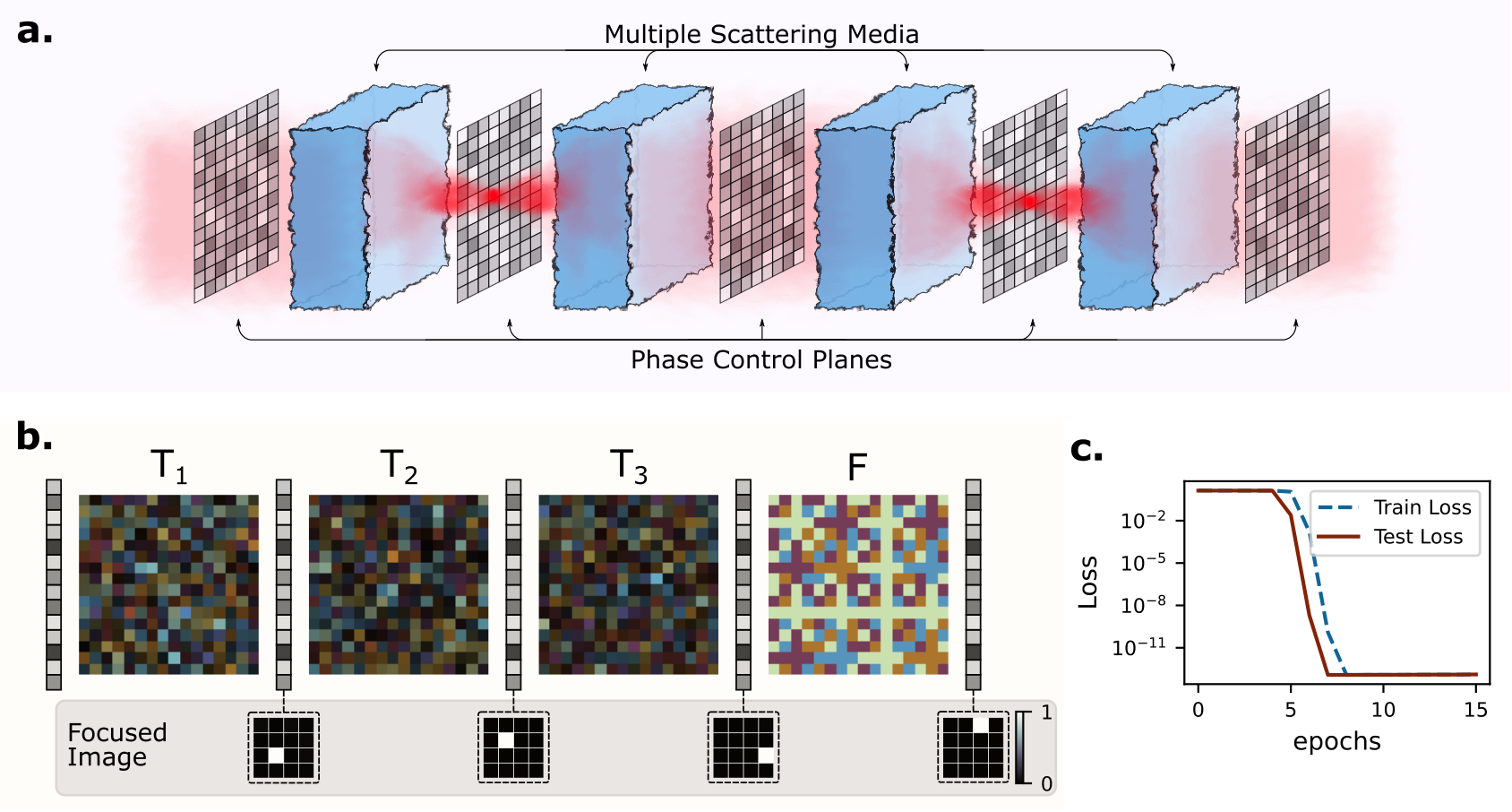}
\caption{Learning multiple transmission matrices with an MPNN: (a) A cascade of multiple complex media interspersed by programmable phase planes can be characterised through the use of the MPNN technique. The recovered TMs corresponding to these media allows us to control the propagation of light at intermediate phase planes. (b) Numerical simulation showing three $16 \times 16$-dimensional TMs reconstructed with the MPNN technique, which are then used to focus light at each intermediate plane (insets). (c) Optimisation loss using the training and testing datasets for the three-plane MPNN.}
 \label{fig:MPNN_results}

\arxivFig{\end{figure*}}
\osaFig{\end{figure}}

%% file: backmatter.tex
\bmsection{Funding}
This work was made possible by financial support from the QuantERA ERA-NET Co-fund (FWF Project I3773-N36), the UK Engineering and Physical Sciences Research Council (EPSRC) (EP/P024114/1) and the European Research Council (ERC) Starting grant PIQUaNT (950402).\\

\bmsection{Acknowledgments}
We would like to thank Dr.~Will McCutcheon and Mayuna Gupta for fruitful discussions. The scientific colour maps batlow and romaO~\cite{crameri_scientific_2021} are used in
this study to prevent visual distortion of the data and exclusion of
readers with colour vision deficiencies~\cite{crameri_misuse_2020}.

\bmsection{Disclosures}
The authors declare no conflicts of interest.


\bmsection{Data availability} Data underlying the results presented in this paper are available in Ref.~\cite{Goel_Data_and_codes_2022}.